\documentclass[twocolumn,showpacs,preprintnumbers,amsmath,amssymb]{revtex4-1}
\usepackage{graphicx}
\usepackage{bm}
\usepackage{float}
\usepackage{color}
\usepackage{ulem}
\usepackage{refstyle}
\usepackage{mathrsfs}
\usepackage{esint}
\usepackage[unicode=true,pdfusetitle,bookmarks=true,
bookmarksnumbered=false,bookmarksopen=false,breaklinks=false,
pdfborder={0 0 0},backref=false,colorlinks=true,citecolor=red]
{hyperref}
\usepackage{cleveref}
\definecolor{dblue}{rgb}{0.0, 0.0, 0.95}

\allowdisplaybreaks[3]

\begin{document}
\title{Coexistence of nonequilibrium phases in assemblies of driven nematic colloids}

\author{Josep M. Pag\`es$^1$}
\author{Arthur V. Straube$^{2,3}$}
\author{Pietro Tierno$^{3,4,5}$}
\author{Jordi Ign\'es-Mullol$^{1,4}$}
\email{jignes@ub.edu}
\author{Francesc Sagu\'es$^{1,4}$}
\affiliation{
$^{1}$Departament de Ci\`encia de Materials i Qu\'{\i}mica F\'{\i}sica, Universitat de Barcelona, Catalonia.\\
$^{2}$Freie Universit\"at Berlin, Department of Mathematics and Computer Science,  Berlin, Germany\\
$^{3}$Departament de F\'{\i}sica de la Mat\`eria Condensada, Universitat de Barcelona, Catalonia.\\
$^{4}$Institut de Nanoci\`encia i Nanotecnologia (IN$^2$UB), Universitat de Barcelona, Catalonia.\\
$^{5}$Universitat de Barcelona Institute of Complex Systems (UBICS), Universitat de Barcelona, Catalonia.
}
\date{\today}
\begin{abstract}
We combine experiments, theory, and simulations to investigate the coexistence of nonequilibrium phases emerging from interacting colloidal particles that are electrokinetically propelled in a nematic liquid crystal solvent.
We directly determine the mechanical pressure within the radial assemblies and measure a non-equilibrium equation of state for this athermal driven system. A generic model combines phoretic propulsion with the interplay between electrostatic effects and liquid-crystal-mediated hydrodynamics, which are effectively cast into a long-range interparticle repulsion, while elasticity plays a subdominant role. Simulations based on this model explain the observed collective organization process and phase coexistence quantitatively. Our colloidal assemblies provide an experimental test-bed to investigate the fundamental role of phoretic pressure in the organization of driven out-of-equilibrium matter.
\end{abstract}
\maketitle


\normalem
\emph{Introduction}. Active colloidal suspensions are a flourishing research field in soft condensed matter, as they
allow investigating the out-of-equilibrium physics of interacting artificial and living entities,
across different length scales~\cite{becquinger2016,zottl2016,elgeti2015}. Given the complexity of these intrinsically non-equilibrium systems, the growing interests to understand their phase behaviour \cite{solon2018} demands alternative approaches that usually extend classical thermodynamic concepts in an effective manner. Particle mobility, which is often considered as an effective temperature, can be used to control aggregation, since activity may lead to the emergence of particle cohesion \cite{palacci2010,ginot2015}.  On the other hand, confinement of active particles may result in an effective pressure exerted by the walls \cite{takatori2014,solon2015,junot2017} and is often related to the local density with a non-equilibrium equation of state, although whether this approach is reliable for generic active systems under confinement is a matter of debate in the community \cite{solon2015}. A rigorous assessment of these issues requires an experimental platform capable to exert control on the size, shape, reversibility, and placement of the aggregates,
which are beyond the capabilities offered by experiments realized so far.

In this article, we report experiments and numerical simulations on the assembly of interacting nematic colloidal swimmers into quasi-two-dimensional aggregates where phase coexistence is driven by a balance between phoretic pressure and long-range repulsion. Unlike conventional studies where colloids are dispersed in isotropic media, our experiments are performed in an anisotropic nematic liquid crystal (LC), which provides both individual and collective control capabilities over the colloidal units \cite{lavrentovich2016,hernandez2013,lazo2014,zhou2014}. In contrast to equilibrium assembly of LC colloids \cite{musevic2017} and in bacteria swimming in a lyotropic LC \cite{zhou2014,peng2016,genkin2017}, here, LC elasticity plays only a minor role, and other properties of the solvent, such as the anisotropy in its ion mobility, enable particle propulsion and pairwise repulsion through LC-enabled electrokinetic phenomena. Force balance 
leads to the coexistence of three colloidal phases in open-boundary assemblies, namely, a dense, solid-like core surrounded by a pressurized liquid-like corona with inhomogeneous density that terminates at an outer diluted gas-like phase. Using direct mechanical arguments, we measure the effective pressure exerted by the driven colloids, which we relate to the steady-state local density. By using numerical simulations, we uncover a rich scenario characterized by a balance between competing mechanisms that arise from variate physical origins and span different length scales. In contrast to existing active particle systems, our driven assemblies provide with a direct way to measure and control the effective pressure, and enable to monitor the formation of a non-equilibrium cluster phase with hexagonal order, unencumbered by bounding walls.

\begin{figure}[t]
  \includegraphics[width=\columnwidth]{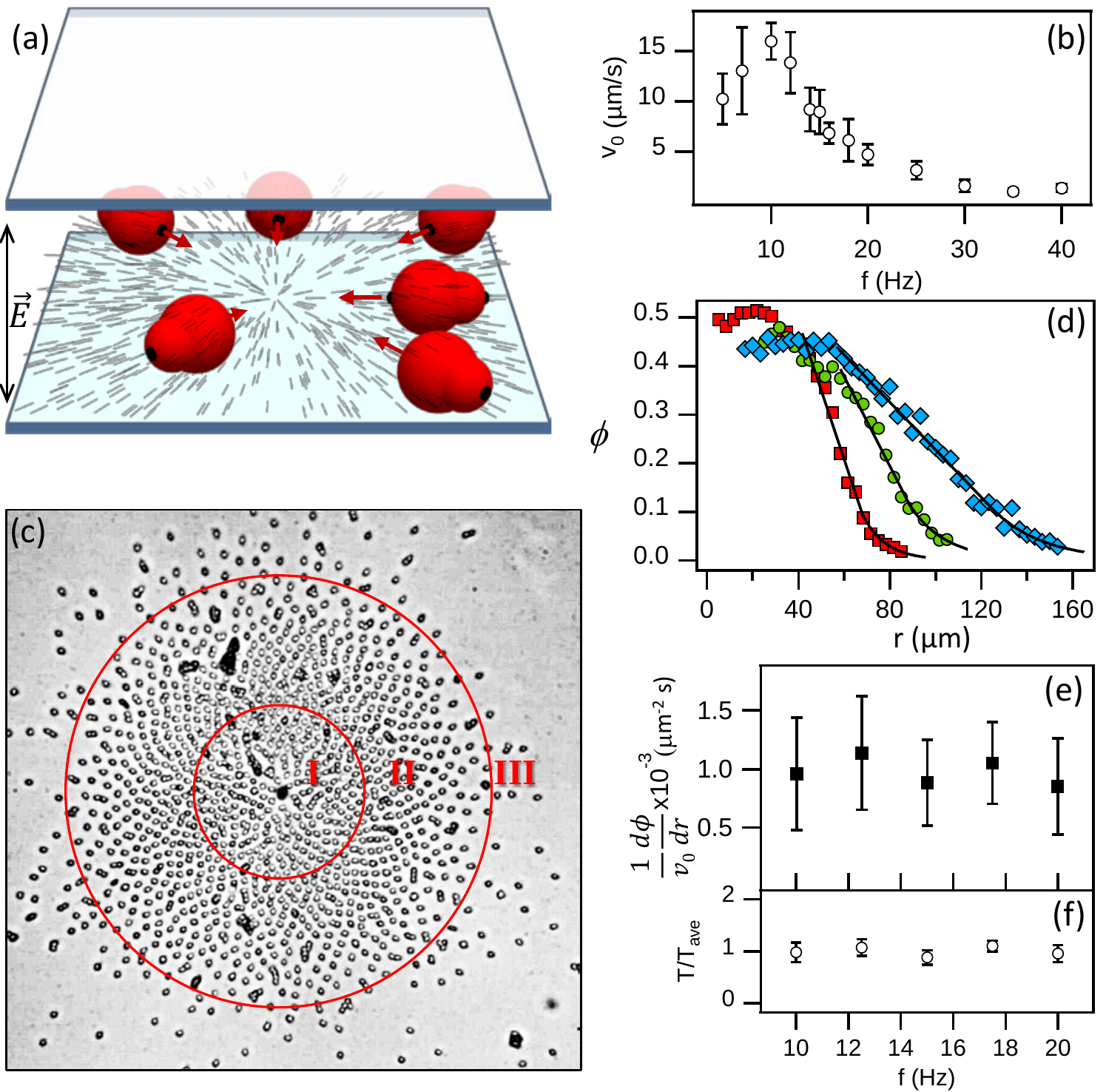}
  \caption{(a) Sketch of the experimental setup. Anisometric colloidal particles dispersed in LC are propelled by an AC electric field normal to the sample plane. (b) Mean velocity of individual particles, $v_0$, as a function of $f$ for $E = 0.76 \rm{V \mu m^{-1}}$. (c) Micrograph of a cluster assembled under a $f=20 \rm{Hz}$ AC field. The field of view is $350\mu$m wide (see also Video S1). (d) Area fraction occupied by the particles {\emph vs} distance from the cluster center for experiments at $f=10\,(\Box)$, $15\,(\circ)$, and $20\,(\diamond)$ Hz. (e) Average slope of the density profile in region (II) normalized by $v_0$, {\emph vs} $f$. (f) Effective temperature relative to its average value, {\emph vs} $f$. }
  \label{fig:assembly}
\end{figure}
%
\emph{Experimental system}. We use polystyrene pear-shaped colloidal particles, $3\times4 \mu$m$^2$ in size (width$\times$length),
dispersed in a nematic LC confined between two parallel transparent electrodes separated by a gap of $20 \,\mu$m. The inclusion shape and the LC are chosen so that a low frequency AC electric field applied between the electrodes leads to an electrokinetic propulsion of the individual particles along the local LC director \cite{lavrentovich2010,hernandez2014,lazo2014}, Fig. \ref{fig:assembly}(a). Because of the viscosity of the medium, the overdamped motion of individual particles occurs with a constant speed $v_0(f)$ (Fig. \ref{fig:assembly}(b)). We restrict our experiments to frequencies $f$ above 10 Hz to ensure a monotonic dependence of the particle speed, and we keep the amplitude of the sinusoidal AC field at $0.76 \rm{V \mu m^{-1}}$. One of the glass plates is coated with a photosensitive layer, which allows to reversibly define a radial alignment of the LC director \cite{hernandez2014,oswald2017}, thus imposing a centripetal driving force $\gamma v_0$ on the particles, where $\gamma$ is the particle friction coefficient. This leads to the dynamic self-assembly of quasi-two-dimensional circular clusters that can grow to arbitrary sizes, Fig.~\ref{fig:assembly}(c).

\emph{Nonequilibrium particle assembly}. In the absence of long-range pairwise repulsion, particles would assemble into compact clusters, as it is commonly observed with isotropic solvents. In the present case, however, we observe the coexistence of three different types of aggregation in the steady-state, emphasizing the existence of long-range repulsion. The innermost core (region I in Fig.\ref{fig:assembly}(c)) is an arrested, jammed state that does not exhibit any further particle rearrangement as the assembly grows. Its average packing density is roughly independent of the distance to the cluster center (Fig. \ref{fig:assembly}(d)), and does not vary significantly for different driving frequencies or, equivalently, particle-level centripetal force.

The intermediate region (II) is a liquid-like state, with particles rearranging over time. It features a particle area fraction, $\phi$, that decreases linearly with the distance to the boundary with region (I) (Fig. \ref{fig:assembly}(d)). As more particles aggregate into the growing cluster, the radial width of region (II) does not change, only the size of the arrested core increases. The spatial arrangement of region II can be understood as the interplay between the phoretic force acting on individual particles, and a net interparticle repulsion force, $F(r)$. Lateral density increases deeper into the assembly because a lower interparticle distance is required for $F(r)$ to balance the accumulated centripetal force exerted by outer particles. Qualitatively, we may interpret the observed linear change of $\phi$ by considering nearest neighbors interaction only and restricting the analysis to the harmonic approximation, which leads to $d\phi/dr\propto - \gamma v_0/U''(q)$. Here, $q$ is the equilibrium hard-core interparticle distance, and $F(r) = -\partial U(r)/\partial r$. Using the measured single-particle speed, we can estimate the scaling of $U(r)$ with frequency from the slope of $\phi(r)$ in region II. Remarkably, we find that the only frequency dependence in the density profile is contained in $v_0(f)$ (Fig. \ref{fig:assembly}(e)), suggesting that the net interparticle repulsion is independent of $f$. As we will show later, this result reveals the rich physics of particle interactions in our system.

Finally, at area fractions below 0.1 (region III), Fig.~\ref{fig:assembly}(c), $\phi(r)$ decays exponentially to zero in analogy to an ideal gas under barometric conditions, and similarly to the reported behavior of sedimenting active colloidal particles \cite{ginot2015}. By considering that a ``body force'' $\gamma v_0$ is exerted upon each particle, we can estimate, for this gas-like region, an effective temperature by fitting $\phi(r) \propto \exp(-\gamma v_0 r/k_B T_{\rm eff})$. As shown in Fig.~\ref{fig:assembly}(f), we find that $T_{\rm eff}$ is frequency-independent, consistently with the athermal nature of our system, and different from other active systems \cite{ginot2015}.

Another remarkable effect of the interplay between directed particle drive and long-range pairwise repulsion is the anomalous evolution of particle ordering within an assembly, which we may quantify with a bond-orientational order parameter. Order is maximum in region II, where centripetal force is balanced by long-range repulsion, it decays to zero towards region III, since density vanishes, but it also decreases towards region I due to the anisotropic short-range interactions. This is in contrast to experiments with sedimenting particles, where the bond-orientational order is a monotonic function of the packing density \cite{thorneywork,ginot2015}.

\emph{Nonequilibrium equation of state}. The different regions observed in our colloidal aggregates are stationary phases formed by propelling particles that constantly exert a radial pressure towards the center of the assembly. Our system is intrinsically out-of-equilibrium and switching off the electric field leads to the melting of the aggregates, as the colloidal particles slowly diffuse along the cell. Thus, we can define a nonequilibrium equation of state that relates the steady-state packing density profiles of the liquid-like region and the computed mechanical pressure that is exerted by the driven particles.
\begin{figure}[t]
	\includegraphics[width=\columnwidth]{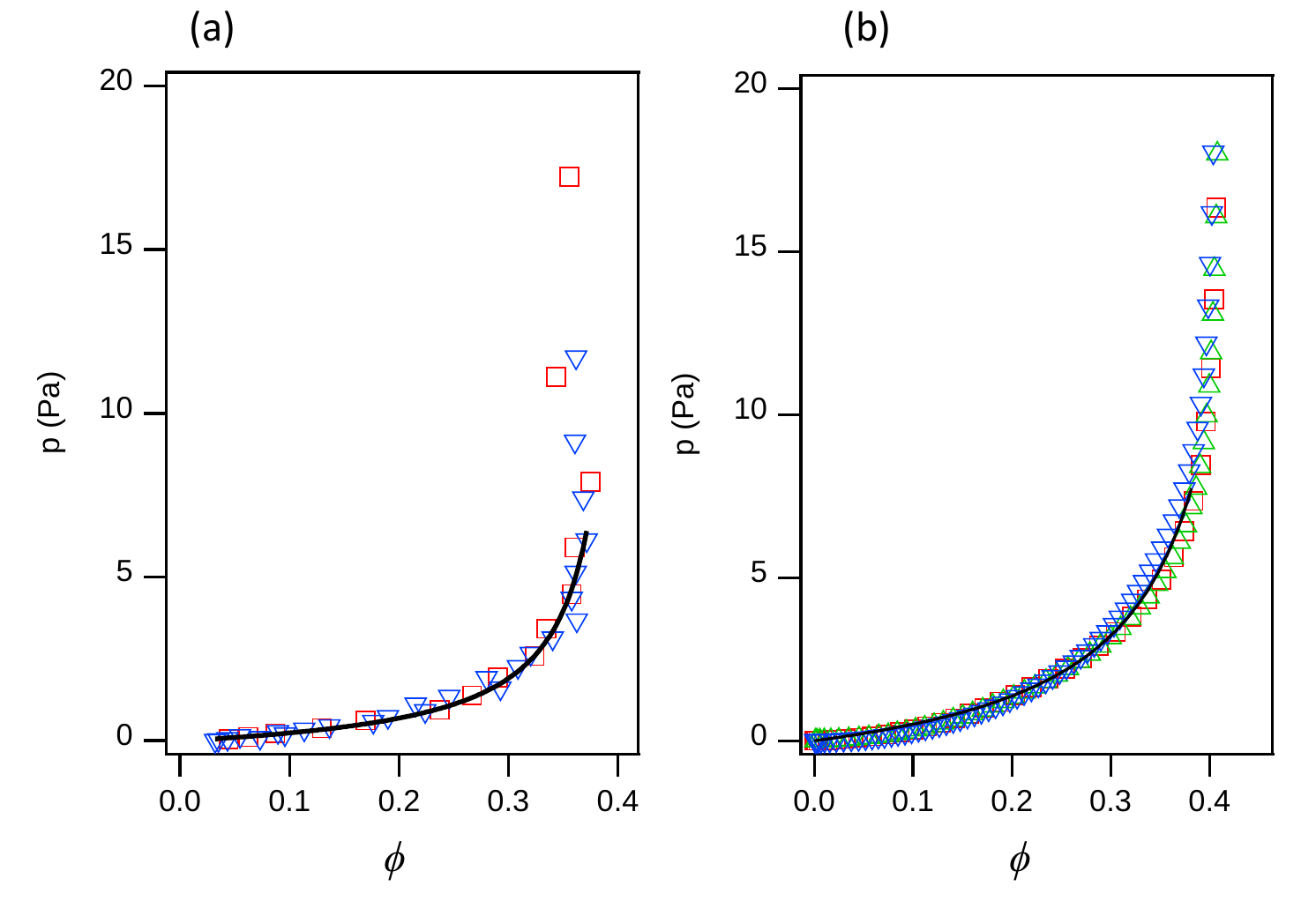}
	\caption{Mechanical pressure as a function of the particle area fraction in an assembled cluster for experiments (a) and simulations (b). The data shown are taken at $10 \rm{Hz}$ ($\Box$), $15 \rm{Hz}$ ($\bigtriangleup$), and $20 \rm{Hz}$ ($\bigtriangledown$). Continuous lines are fits to a hard-disk equation of state in the compressible, liquid-like region.
	}
	\label{fig:thermo}
\end{figure}
We define the mechanical pressure $p$ such that it is zero outside of the cluster, and its value at a distance $r$ from the cluster center is given by
\begin{equation}
p(r)=\frac{F}{w\,r}\int_{r}^{\infty}\frac{\phi(r')}{a_0}r'dr'  \, \, .
\label{eq:pressure}
\end{equation}
Here, $a_0 \simeq 10 \mu\rm{m}^2$ is the area covered by a single particle, and $w$ is the thickness of the particle layer, which we take as the diameter of the largest particle lobule, $w=3 \rm{\mu m}$.
As a first approximation, we will assume that the force $F$ is the same that propels isolated particles, $F=\gamma v_0(f)$, and consider that friction is solely due to the solvent viscosity, thus neglecting the effect of substrate friction. Consequently, we estimate $\gamma = 6\pi\eta_{\parallel} a$, where  $a=1.5 \rm{\mu m}$ is the particle hydrodynamic radius and $\eta_{\parallel}=1.4\times 10^{-3}$ Pa s \footnote{We measured the reported value in a separate experiment, by comparing the motion of paramagnetic colloidal particles propelled by an external magnetic field gradient both in water and in the aligned LC.} is the LC viscosity along the local director.

The results of this data analysis are presented in Fig. \ref{fig:thermo} as $p$-$\phi$ isotherms, which are continuous through all the aggregation stages. Consistently with the experimental observation that particle interactions are effectively frequency-independent, our results show that the isotherms obtained under different driving conditions collapse onto a single master curve. Moreover, unlike other aggregation experiments reported with active \cite{ginot2015} or driven passive diffusive micro-particles \cite{thorneywork}, thermal fluctuations are negligible in our system, given the relatively large viscosity of the dispersing medium.
The observed trend for the $p$-$\phi$ data in region II of the colloidal assemblies can be well described by an equation of state similar to the hard-disk case \cite{santos95},
\begin{equation}
p\propto \frac{\phi}{1-2\phi + (\phi_0^2-1)\left(\frac{\phi}{\phi_0}\right)^2},
\end{equation}
where $\phi_0$ is the extrapolated close-packing particle area fraction. Such behavior may be understood by considering that, under the diluted conditions of region II, the large interparticle distances prevent the LC-mediated elastic interactions to set in, thus rendering the colloidal behavior effectively similar to a hard-sphere liquid.
\begin{figure}[!t]
	\includegraphics[width=\columnwidth]{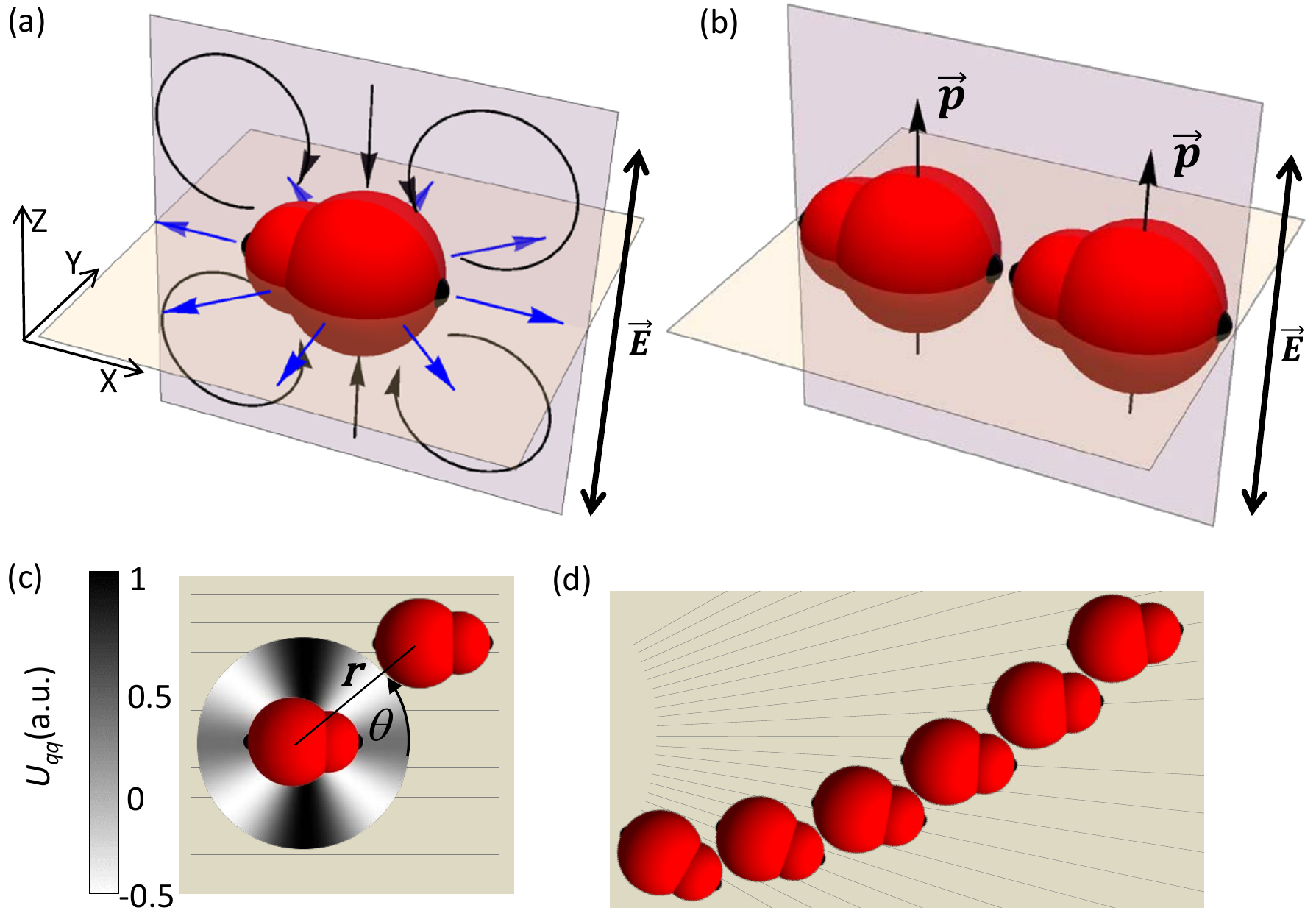}
	\caption{Sketch of the different interparticle interactions. (a) Hydrodynamic repulsion due to phoretic ion flows, which are pushed away from the particles in the plane of the assembly (blue arrows). The local director field is along the X axis. (b) Induced electric dipoles are parallel, co-planar, and perpendicular to the plane of the assembly, resulting in a net repulsion. (c) LC-mediated elastic quadrupolar interactions, $U_{qq}$, are anisotropic, and become attractive when the long axes of particles are shifted about $45^{\circ}$ with respect to the local LC director. (d) Because of the radial LC director, elasticity promotes the formation of arched particle chains.
	}
	\label{fig:interact}
\end{figure}

\emph{Theoretical model}.
To understand the observed colloidal assembly process we performed numerical simulations for a layer of $N$ hard-spheres confined in a plane and driven along the local orientation of a LC director field. In analogy with the experiments, we consider that the LC director features a radial pattern emanating from a center. We assume that the particles are propelled at a constant, frequency-dependent speed $v_0(f)$ arising from LC-enabled electrokinetic drive \cite{lavrentovich2010}.
Interaction with neighboring particles modulates the instantaneous particle velocity, as described by the overdamped equations
\begin{align}
\frac{d\bm{r}_i}{dt} & = v_{0,i}(f)\hat{\bm{n}}(\bm{r}_i)-\frac{1}{\gamma}\sum_{j\ne i}\frac{\partial U }{\partial\bm{r}_i}. \label{eom}
\end{align}
Here, $\bm{r}_i=(x_i,y_i)$ is the position of particle $i$, $\hat{\bm{n}}(\bm{r})$ is the local LC director, $\gamma$ is the friction coefficient, and $U$ is the pairwise interaction potential,
\begin{align}
U & = U_{hd}(r_{ij}) + U_{dd}(r_{ij}) + U_{hc}(r_{ij}) + U_{qq}(\bm{r}_{ij},\hat{\bm{n}}), \label{pot-U}
\end{align}
where $U_{hd}$ is a long-range hydrodynamic repulsion, $U_{dd}$ is an induced dipole repulsion,
 $U_{hc}$ is a short-range hard-core repulsion, and $U_{qq}$ is the short-range elastic quadrupolar interaction mediated by the LC matrix. 

\emph{Hydrodynamically induced repulsion}. Experimental studies by Lazo \emph{et al}\cite{lazo2014} revealed that, under an applied AC field, particles with planar LC anchoring push ionic flows away from them. While their near-field determines the colloidal self-propulsion\cite{lavrentovich2010}, their far-field mediates a hydrodynamic coupling between neighboring particles. In consequence, $v_0$ and $U_{hd}$ share the same frequency dependence, $\alpha(\omega)$, where $\omega = 2\pi f$. Moreover, in our experiments, the electric field is perpendicular to the plane of particle motion. Thus we conclude that induced ionic flows move radially outwards from each particle in the plane of the assembly, independently of the in-plane direction (Fig. \ref{fig:interact}(a)). Finally, we consider the known $r^{-3}$ far-field decay of hydrodynamic flows in cell geometries  \cite{lazo2014,peng2014}, resulting in a repulsive potential
\begin{align}
U_{hd}(r,\omega) = C_{hd} \frac{\alpha(\omega)}{r^2}, \alpha(\omega) = \frac{ (\tau_s+\tau_e)^2\omega^2 }{ (1 + \tau_s^2\omega^2) (1 + \tau_e^2\omega^2)},
\label{U-hd}
\end{align}
with a constant prefactor $C_{hd}$ that will be fitted to the experimental data.

\emph{Induced electrostatic dipolar repulsion}. Although the repulsive potential $U_{hd}$ would be enough in the model to balance the phoretic drive and to reproduce the linear density profile, it fails to capture the experimentally observed scaling of $\phi(r)$ with $v_0(f)$ (Fig. \ref{fig:assembly}), since the frequency dependencies of $U_{hd}$ and $v_0$ are the same. Because of this, we need to invoke the induced dipole moments $\bm{p}$ that appear as a result of charge separation leading to the formation of electric double layers surrounding each particle (Fig. \ref{fig:interact}(b)). In fact, alternative geometries allow to put into evidence the strength of electrostatic interactions, which may dominate particle assembly in other arrangements. The resulting pairwise interaction can be expressed as $U_{dd}(r,\omega) \propto {\bm p}^2/r^3$ which, using the known frequency dependence of these induced dipoles \cite{squires2004,murtsovkin96}, leads to
\begin{align} \label{U-dd}
U_{dd}(r,\omega) = C_{dd} \frac{\beta(\omega)}{r^3}, \quad
 \beta(\omega)=\frac{(1/4+\tau_s^2\omega^2) \tau_e^2 \omega^2}
{(1+\tau_s^2\omega^2)(1+\tau_e^2\omega^2)},
\end{align}
where the frequency-independent prefactor $C_{dd}$ will be fitted to the data. Note that $U_{hd}$ and $U_{dd}$ have complementary frequency dependencies, which is crucial to reproduce the experimentally observed behavior.

\begin{figure}[t!]
  \includegraphics[width=\columnwidth]{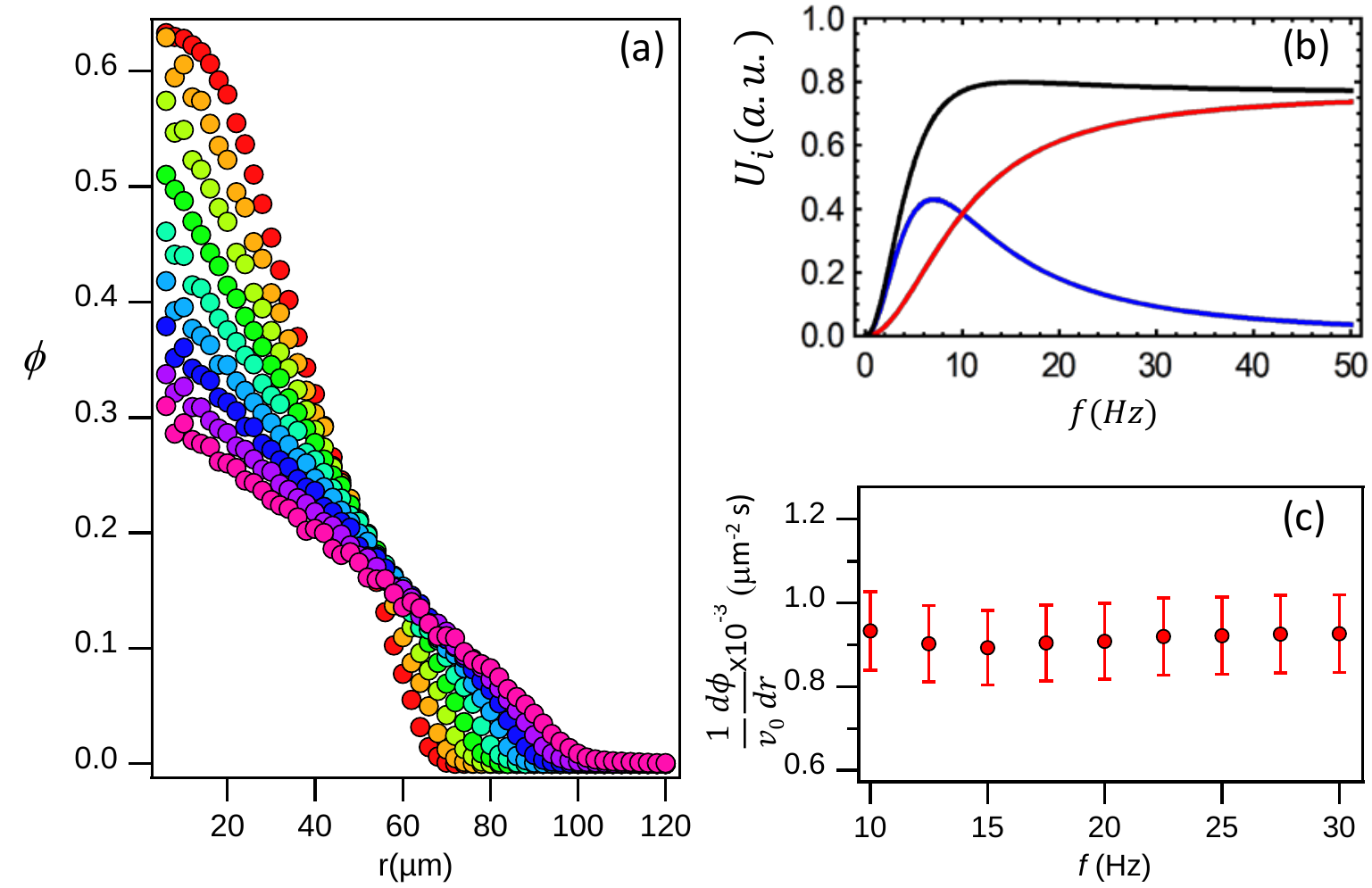}
  \caption{(a) Area fraction occupied by particles as a function of distance from the center of the cluster for simulated cluster assemblies with different driving frequencies (see also Video S1). The data corresponds to clusters with $700$ particles. From top to bottom, frequencies range from $10\rm{Hz}$ to $30\rm{Hz}$ in steps of $2.5\rm{Hz}$. (b) Hydrodynamic ($U_{hd}$, blue), induced-dipolar ($U_{dd}$, red), and combined ($U_{hd}+U_{dd}$, black) repulsion potentials, calculated for a center-to-center distance of $15 \rm{\mu m}$.(c) Slope of the linear region of the density profiles in (c), normalized by the phoretic speed, {\emph vs} driving frequency.
  }
  \label{fig:sim_slopes}
\end{figure}

\emph{Short-range interactions}. At short distances within region I, hard core repulsion and LC-mediated elasticity dominate, both frequency-independent interactions. The former is introduced by means of $U_{hc}(r)$, whose exact form has little influence on the assembly parameters, while the latter takes the anisotropic form
\begin{equation}\label{U-qq}
U_{qq}(\bm{r},\hat{\bm{n}})=\frac{C_{qq}}{r^5}(3-30 \cos^2 \vartheta +35 \cos^4 \vartheta).
\end{equation}
Here, $\vartheta=\vartheta(\bm{r},\hat{\bm{n}})$ is the angle between the vector connecting the centers of particles, $\bm{r}$, and the far-field orientation of the nematic director (Fig. \ref{fig:interact}(c)). In the denser parts of the cluster, the quadrupolar LC structure around the particles promotes their off-centered chaining \cite{smalyukh2005,musevic2017}, leading to the formation of arch-like chains (Fig. \ref{fig:interact}(d)).

\emph{Choice of elektrokinetic parameters}. From the experimental speed of individual particles, we can fit the intrinsic time scales that characterize the long-range repulsion terms, yielding $\tau_s \approx 0.016 \,{\rm s}$ and $ \tau_e \approx 0.032 \,{\rm s}$. In order to reproduce the experimental observations, we use in our simulations $C_{hd}/\gamma = 29\, {\rm \mu m^4/s}$ and $C_{dd}/\gamma = 517 \, {\rm \mu m^5/s}$ (Fig. \ref{fig:sim_slopes}). With this, we find that, as observed experimentally, the combined potential, $U_{hd}+U_{dd}$, is nearly frequency-independent in the used range of frequencies (Fig. \ref{fig:sim_slopes}(b)). Finally, we set $C_{qq}/\gamma=150 \, \rm{\mu m^7}/s$ to fine tune the agreement with experiments (Fig. \ref{fig:sim_slopes}(c)). We have included 20\% Gaussian noise in the individual speed of simulated particles in analogy with experimental observations.  Note that, for the number of particles in the simulated ensembles in Fig. \ref{fig:sim_slopes}, the formation of a solid-like core is found only at $f=10 \, \rm{Hz}$. For a larger number of particles, the core can be observed at all driving frequencies.

\emph{Conclusions}. We have investigated the assembly of driven colloidal particles in a nematic liquid crystal, whose anisotropy, combined with the geometry of the experiments, result in the coexistence of three colloidal phases. Numerical simulations reveal that the observed assemblies result from the balance between a disparity of electrokinetic phenomena involving hydrodynamics, phoretic forces, and dipolar interactions.
Quantitative agreement between the experiment and model  indicates that all particular ingredients of the system, including complex hydrodynamic flows generated around particles, can be cast into a general framework of individual propulsion and long-range  pairwise repulsive forces. Importantly, the elasticity of the liquid crystal plays only a secondary role in the observation of phase coexistence, for which the choice of a radial geometry is not essential, and similar effects can be expected for a parallel geometry.
On the whole, our experiments demonstrate enhanced control capabilities for microscale colloidal assembly endowed by their dispersion in anisotropic solvents.

\begin{acknowledgments}
We thank A. Ortiz-Ambriz for assistance in measuring the LC viscosity. The authors acknowledge fruitful discussions with L. Schimansky-Geier and I. Sokolov. J. M. P. acknowledges funding from
the European Union's Horizon 2020 Fetopen "AbioMatter",
grant No. 665440. Experiments were funded by MINECO ( AEI/FEDER, EU), projects FIS 2013-41144P and FIS2016-78507-C2-1-P .
P.T. acknowledges support from the European Research Council grant agreement No. 335040, from MINECO ( AEI/FEDER, EU), project FIS2016-78507-C2-2-P, and DURSI, project 2014-SGR-878. A. V. S. acknowledges partial support from the Excellence Initiative and SFB1114 of the German Research Foundation.
\end{acknowledgments}
%

\end{document}